\documentclass[12pt]{article}
\usepackage{amssymb}
\begin{document}

\begin{titlepage}

                            \begin{center}
                            \vspace*{2cm}
        \large\bf The geodesic rule and the spectrum of the vacuum.\\

                            \vfill

              \normalsize\sf    NIKOS \ \ KALOGEROPOULOS\\

                            \vspace{0.2cm}

 \normalsize\sf Department of Science\\
 BMCC - The City University of New York,\\
 199 Chambers St., \ New York, NY 10007, \ USA\\

                            \end{center}

                            \vfill

                     \centerline{\normalsize\bf Abstract}
\normalsize\rm\setlength{\baselineskip}{18pt} \noindent We analyze
the consequences of a recent argument justifying the validity of
the ``geodesic rule" which can be used to determine the density of
global topological defects. We derive a formula that provides a
rough estimate of the number of string-like defects formed in a
phase transition. We apply this formula to vacua which are
spheres. We provide some reasons for the deviation of our
predictions from the corresponding accepted values.

                             \vfill

\noindent\sf PACS: \ \ \ \ \ \ 02.50.Ey, 11.10.Ef, 11.30.Fs\\
    Keywords: \ Kibble-Zurek mechanism, Geodesic rule, Defect density, Spectrum of the Laplacian.\\

                             \vfill

\noindent\rule{8cm}{0.2mm}\\
\begin{tabular}{ll}
\small\rm E-mail: & \small\rm nkalogeropoulos@bmcc.cuny.edu\\
                  & \small\rm nkaloger@yahoo.com
\end{tabular}
\end{titlepage}


                            \newpage

\normalsize\rm\setlength{\baselineskip}{18pt}

 The geodesic rule [1] is a way that allows us to develop predicting
 schemes for the density
 of global topological defects arising in phase transitions [2],[3].
 Two approaches exist for justifying the geodesic rule.
 The first, time-honored  and  widely accepted,
 relies on the minimization of the energy of the system [4],[5].
 We  recently proposed [6] a second approach which relies
 on the Markovian character of the dynamics and on the random
 orientation of the collisions  among the expanding causally connected
 volumes of true vacuum. Such collisions take place inside the meta-stable
 vacuum background. In the present paper we follow the second approach,
 and  propose a formula that provides a rough estimate for the number
 of global topological strings formed in a phase transition.\\

 Topological strings are of great importance methodologically [2],[3] due to
 their non-perturbative nature. What is even more important, they
 have severe experimental and observational implications [2],[3],[7] in models
 allowing their formation. Because of their great significance, a scheme leading
 to calculations of their density and distribution
 such as the geodesic rule has to be theoretically justified and its predictions
 have to be thoroughly tested. Sometime after [1], the importance of the role
 of the temperature quench and the effects of the related critical slowing down
 of the underlying phase transition
 were pointed out in [8]-[11].
 The predictions relying on this modification of the geodesic
 rule, the Kibble-Zurek scenario,
 are generally considered to be in agreement with experimental
 data [12]-[18] and numerical simulations [19]-[23].
 Any discrepancies between such experimental/numerical data and the theoretical
 predictions are attributed to the details of the dynamics of the particular
 model under study, rather than being considered a shortcoming of the general approach.
 Despite all these successes, there is still some work that
 needs to be done, especially on the analytical side, in finding formulae
 expressing the density of topological defects. The current study,
 aims in such a direction.\\

 Let \ $\mathcal{M}$ \ denote the vacuum of the model
 under study and \ $\phi \in \mathcal{M}$ \ be a parametrization of it.
 We arbitrarily pick \ $\phi=0$ \  on \ $\mathcal{M}$ \
 as the origin of a normal coordinate system on which
 the calculations rely. As was explained in [6], the stochasticity of the
 orientation of the causally connected volumes during their collisions and the
 Markovian character of their subsequent coalescence can be
 described by a probability distribution function \ $P(\phi,t)$ \
  which obeys the Fokker-Planck equation
\begin{equation}
   \frac{\partial P(\phi, t)}{\partial t} \ = \ D \nabla^2 P(\phi, t)
\end{equation}
Here \ $D$ \ stands for the diffusion constant of this stochastic
process on \ $\mathcal{M}$ [6]. \ The solution of (1), with the
initial condition
\begin{equation}
 P(\phi,0)=\delta(\phi)
\end{equation}
is the heat kernel \ $K(0,\phi,t):
\mathcal{M}\times\mathcal{M}\times\mathbb{R}_+ \rightarrow
\mathbb{R}$. \ We recall that \ $K(0,\phi,t)$ \ has the
eigenfunction expansion
\begin{equation}
 K(0,\phi,t)=\sum_{j=0}^{\infty}e^{-\lambda_jt}f^{\ast}_j(0)f_j(\phi)
\end{equation}
where $\ast$ denotes complex conjugation, $\lambda_j,
j=0,1,\ldots$ are the eigenvalues of \ $\nabla^2$ \ on \
$\mathcal{M}$. Here \ $f_j(\phi):\mathcal{M}\rightarrow\mathbb{R}$
\ denote the normalized eigenfunctions of \ $\nabla^2$ \ on
$\mathcal{M}$ \ with respect to the hermitian product
\begin{equation}
      (f_j, f_l) =  \int\limits_{\mathcal{M}} f^{\ast}_j(\phi)f_l(\phi)d\phi =
         \delta_{jl}, \ \ j,l = 0,1,\ldots
\end{equation}
induced by the Riemannian metric \ $g$ \ of \ $\mathcal{M}$  \ on
the space of the square integrable functions \ $L^2(\mathcal{M})$.
\ We assume, for simplicity, that the vacuum \ $\mathcal{M}$ \ is
a manifold without boundary. Since \ $\nabla^2$ \ is an elliptic,
positive semi-definite operator on \ $\mathcal{M}$, then
$\lambda_j\geq 0, \ j=0,1,\ldots$. Moreover, since \ $\nabla^2$ \
is self-adjoint with respect to the inner product \ $(\cdot
,\cdot)$, \ then \ $f_j(\phi)\in\mathbb{R}, \ j=0,1,\ldots$.\\

We also assume, for concreteness, that the phase transition giving
rise to the topological strings is of first order and proceeds by
bubble nucleation [6]. Given this assumption, consider three such
bubbles with corresponding values of the order parameter \
$\phi_1=0$, \ $\phi_2$, \ $phi_3$. \ When these bubbles collide
the geodesic rule states that \ $\phi$ \ traverses a piecewise
smooth loop \ $C$ \ on \ $\mathcal{M}$. \ Without any loss of
generality assume \ $C$ \ has base point \ $\phi_1=0$. \ The
points on \ $\mathcal{M}$ \ at which \ $C$ \ is non-smooth, with
probability almost 1, are exactly \ $0, \ \phi_2, \ \phi_3$. \ If
\ $C$ \ is non-contractible then the collision of the three
bubbles  will give rise to a topological defect. The fact that to
determine the formation or not of a topological string one must
follow a piecewise smooth geodesic loop \ $C$, \ shows that the
quantity most closely related to such a string formation is not
the generic heat kernel \ $K(0,\phi,t)$ \ but its diagonal, namely
\ $K(0,0,t)$. \ This can be expressed as
\begin{equation}
 K(0,0,t)=\sum_{j=0}^{\infty}e^{-\lambda_jt}f_j(0)f_j(0)
\end{equation}
Since all points \ $\phi\in\mathcal{M}$ \ are equally likely to be
the base-point of  \ $C$, \ we have to use in what follows instead
of \ $K(0,0,t)$ \ its average over \ $\mathcal{M}$. \ Such an
average is called the partition function \ $Z_{(\mathcal{M},g)}$,
\ and is defined as [24]
\begin{equation}
  Z_{(\mathcal{M},g)}(t) \ := \ \frac{1}{Vol_g(\mathcal{M})}
        \sum_{j=0}^{\infty}e^{-\lambda_jt}\int_{\mathcal{M}}f_j(\phi)f_j(\phi)d\phi
\end{equation}
In (6), \ $Vol_g(\mathcal{M})$ \ is the volume of \ $\mathcal{M}$
\ with respect to the Riemannian measure induced by \ $g$. \ Since
the eigenfunctions $f_j(\phi), \ j=0,1,\ldots$ are orthonormal
with respect to \ $(\cdot , \cdot)$, \ (6) simplifies to
\begin{equation}
 Z_{(\mathcal{M},g)}(t) \ = \ \frac{1}{Vol_g(\mathcal{M})}
                        \sum_{j=0}^{\infty}e^{-\lambda_jt}
\end{equation}
An explicit calculation of \ $Z_{(\mathcal{M},g)}(t)$ \ amounts to
determination of the spectrum \ $\mathrm{Spec} (\mathcal{M}) := \{
\lambda_j, j=0,1,\ldots \} $ \  of the  Laplace-Beltrami operator
\ $\nabla^2$ \ on functions of \ $\mathcal{M}$. \ Conversely, one
can prove [24] that knowing \ $Z_{(\mathcal{M},g)}(t)$ \ amounts
to determining \ $\mathrm{Spec}(\mathcal{M})$. \ Unfortunately, it
is not known how to explicitly calculate \
$\mathrm{Spec}(\mathcal{M})$ \ or equivalently \
$Z_{(\mathcal{M},g)}(t)$ \ in general, except for very few cases
of manifolds with very high symmetry as well as some of their
quotients. It is known, for example [24], that the $n$-sphere \
$\mathbb{S}^n$ \ endowed with its round metric \ $g$, \ has
\begin{equation}
  \mathrm{Spec}(\mathbb{S}^n) \ = \ \{\lambda_k = k(n+k-1), \ \ \ k\in\mathbb{N} \}
\end{equation}
with corresponding multiplicities
\begin{equation}
   m_k  \ = \ \frac{(n+k-2) !}{(n-1)! \ k!}  \ (n+2k-1), \ \ k\geq 1
   \hspace{5mm}  \mathrm{and} \hspace{5mm} m_0 = 1
\end{equation}
Since [24]
\begin{equation}
   Vol_g(\mathbb{S}^n) \ = \ \frac{2\pi^{\frac{n+1}{2}}}{\Gamma
   (\frac{n+1}{2})}
\end{equation}
we find
\begin{equation}
  Z_{(\mathbb{S}^n, g)}(t) \ = \ \frac{\Gamma (\frac{n+1}{2})}{2\pi^{\frac{n+1}{2}}}
       \left\{ 1+ \sum_{k=1}^{\infty} \frac{(n+k-2) !}{(n-1)! \ k!}  \ (n+2k-1)
        \ e^{-k(n+k-1) t} \right\}
\end{equation}
Other examples of spaces that are occasionally found as  vacua of
classical field theories, for which the spectrum is explicitly
known [24] are the real and complex projective spaces \
$\mathbb{R}P^n$ \ and \ $\mathbb{C}P^n$ \ respectively, the tori \
$\mathbb{T}^n$, \ the lens spaces \ $L(p,q)$, \
and somewhat more rarely, the Heisenberg manifolds. \\

Major problems exist in the calculation of the spectrum if \
$(\mathcal{M}, g)$ \ happens to have non-positive sectional
curvature everywhere. Then, according to the Hadamard-Cartan
theorem [25], the universal cover of \ $\mathcal{M}$ \ is
diffeomorphic to \ $\mathbb{R}^{\dim\mathcal{M}}$ \ from which we
cannot draw enough useful information that can help determine \
$\mathrm{Spec}(\mathcal{M})$. \ In most cases of physical interest
though, we do not have to worry about manifolds of non-positive
sectional curvature. Indeed, most  frequently, the
classical/non-thermal vacua \ $\mathcal{M}$ \ happen to be
compact, semi-simple Lie groups \ $G$, \ or their homogeneous
spaces endowed with the corresponding Killing-Cartan metrics or
their images under Riemannian submersions [25]. These spaces have
non-negative sectional curvature everywhere [26], so there is no
need to worry about the implications of
the Hadamard-Cartan theorem in future considerations.\\

In most other cases we have to settle with considerably less;
within the scope of the heat-kernel methods employed in this work,
the short-time behavior of the trace of the heat kernel is
probably the most useful approximation. As the word states, such
an asymptotic expansion is employed when \ $ t\rightarrow 0^+ $ \
and states that
\begin{equation}
 K(\phi, \phi, t) \ \sim \ (4\pi t)^{-\frac{\dim
 {\mathcal{M}}}{2}} \sum_{i=0}^{\infty} u_i(\phi, \phi )t^i
\end{equation}
where \ $u_i(\phi ,\phi)$ \ are polynomials involving the Riemann
tensor, its covariant derivatives and their contractions at \
$\phi\in\mathcal{M}$. \ Substituting (12) into (7), we find the
corresponding asymptotic expansion of
\begin{equation}
Z_{(\mathcal{M},g)}(t) \ \sim \ \sum_{i=0}^{\infty} a_i
t^{\frac{i-dim\mathcal{M}}{2}}
\end{equation}
where
\begin{equation}
a_{2i} \ = \ \frac{1}{(4\pi)^{\frac{dim{\mathcal{M}}}{2}}}
   \int\limits_{\mathcal{M}} u_i(\phi, \phi) \sqrt{\det g} \ d\phi
      \ \ \ \ \ \ \ i\in\mathbb{N}
\end{equation}
and
\begin{equation}
a_{2i+1} \ = \ 0, \ \ \ \ \ \ \ i\in\mathbb{N}
\end{equation}
In particular,
\begin{equation}
 a_0 \ = \ \frac{1}{(4\pi)^{\frac{dim\mathcal{M}}{2}}} \ Vol_g(\mathcal{M})
\end{equation}
\begin{equation}
 a_2 \ = \ \frac{1}{6(4\pi)^{\frac{dim\mathcal{M}}{2}}}
 \int\limits_{\mathcal{M}} \mathrm{R}(\phi) \sqrt{\det g} \ d\phi
\end{equation}
\begin{equation}
 a_4 \ = \ \frac{1}{360(4\pi)^{\frac{dim\mathcal{M}}{2}}}
 \int\limits_{\mathcal{M}}  \left\{ 2||\mathrm{Riem}(\phi)||^2 -
 2||\mathrm{Ric}(\phi)||^2 + 5 \mathrm{R}^2(\phi) \right\} \sqrt{\det g} \ d\phi
\end{equation}
where \ $||\cdot ||$ \ denotes the point-wise norm
\begin{equation}
   ||\mathrm{Riem}(\phi)||^2 \ = \ R_{klpq}R^{klpq}, \ \ \ \ \
              ||\mathrm{Ric}(\phi)||^2 \ = \ R_{pq}R^{pq}
\end{equation}
of the Riemann and the Ricci tensors respectively and \
$\mathrm{R}$ \ denotes the Ricci scalar. The coefficient \ $a_6$ \
is also known [24], but its expression is too long to state and
not very enlightening. The zeroth order term has a clear geometric
interpretation as a multiple of \ $Vol_g(M)$. \ The second-order
term is proportional to the Euler characteristic
 \ $\chi(\mathcal{M})$ \ when \ $\mathcal{M}$ \ is a Riemann surface.
 Unfortunately, in all other cases the non-trivial higher order coefficients
 \ $a_i$ \ do not have a straightforward geometric or topological
 interpretation. So despite of their usefulness in this asymptotic expansion,
 they are not of any great help in improving our intuition
 about the behavior of the diagonal of the heat kernel. Subsequently their
 potential physical significance is also relatively limited, so we will not
 expand further upon them.\\

During consecutive bubble collisions, piece-wise smooth geodesics
of all possible lengths on \ $\mathcal{M}$ \ describe the
evolution of \ $\phi$. \ A more appropriate expression related to
the density of topological defects is then
\begin{equation}
         \mathcal{Z}_{(\mathcal{M},g)} \ = \ \int\limits_0^{+\infty} \
         Z_{(\mathcal{M},g)}(t) \ dt
\end{equation}
Substituting (7) into (20), we find
\begin{equation}
   \mathcal{Z}_{(\mathcal{M},g)} \ = \ \frac{1}{Vol_g(\mathcal{M})}
      \left\{ \int\limits_0^{+\infty} e^{-\lambda_0t}dt +
      \sum_{i=1}^{\infty} \frac{1}{\lambda_i} \right\}
\end{equation}
In (21), the first term gives an infinite contribution since for a
closed manifold like \ $\mathcal{M}$ \ the lowest eigenvalue of
the Laplace-Beltrami operator is zero \ $\lambda_0=0$. \ An
infinite subtraction of \ $\mathcal{Z}_{(\mathcal{M}, g)}$ \ is
required to make sense out of (21). The second term of (21) can be
recognized as the value at \ $t=1$ \ of the zeta function \
$\zeta_{\nabla}(t)$ \ associated with \ $\nabla^2$, \ so we can
write
\begin{equation}
 \mathcal{Z}_{(\mathcal{M},g)} \ = \ \frac{1}{Vol_g(\mathcal{M})}
      \left\{ \zeta_{\nabla}(1) + \int\limits_0^{+\infty} e^{-\lambda_0t}dt
      \right\}
\end{equation}
The function \ $\zeta_{\nabla}(t)$ \ is meromorphic  over \
$\mathbb{C}$ \ with isolated simple poles at \
$\frac{dim\mathcal{M}-j}{2}\in\mathbb{R}$ \ with \
$j=0,1,2,\ldots$ \ and corresponding residues
\begin{equation}
     \mathrm{Res}_{j=j_0} = \frac{a_{j_0}}{\Gamma \left( \frac{dim\mathcal{M} -
     j_0}{2} \right)}
\end{equation}
where \ $\Gamma(z), \ z\in\mathbb{C}\backslash \{ 0, -1, -2,
\ldots \}$ \ denotes the Euler gamma function. If  \
$dim\mathcal{M}$ \ is large enough so that  \
$j_0=dim\mathcal{M}-2 \in\mathbb{N}$ \ namely if \
$dim\mathcal{M}\geq 3$ \ then \ $t=1$ \ is a simple pole of \
$\zeta_{\nabla}(t)$ \ with residue \ $a_{dim\mathcal{M}-2}$. \ In
case \ $dim\mathcal{M}$ \ is an odd number, then the residue at \
$t=1$ \ is zero in accordance with the general result stated
above. If a vacuum for which \ $dim\mathcal{M}\geq 4$ \ happens to
also satisfy \ $a_{dim\mathcal{M}-2} = -1$, \ and if in addition
we interpret the improper integral in (22) as the  weak limit
($\lambda_0=0$)
\begin{equation}
  \int\limits_0^{+\infty} e^{-\lambda_0 t}dt \ = \
  \lim_{\epsilon\rightarrow 0} \int\limits_0^{+\infty} e^{-\epsilon
  t} dt
\end{equation}
then the two infinities of (22) cancel exactly each other and the
final result for \ $\mathcal{Z}_{(\mathcal{M},g)}$  \ is finite.
In a generic case, however, an infinite subtraction from  \
$\mathcal{Z}_{(\mathcal{M},g)}$ \ is required. For the case of
spheres \ $\mathbb{S}^n$, \ (11)
\begin{equation}
 \mathcal{Z}_{(\mathbb{S}^n, g)} \ =
       \ \frac{\Gamma (\frac{n+1}{2})}{2\pi^{\frac{n+1}{2}}}
       \left\{\sum_{k=1}^{\infty} \frac{(n+k-2) !}{(n-1)! \ k!} \cdot
       \frac{(n+2k-1)}{k(n+k-1)} \ +
         \int\limits_0^{\infty} e^{-\lambda_0 t} dt  \right\}
\end{equation}
We can check that the integrated partition functions of (25) are,
with few exceptions, divergent even if we disregard the term
corresponding to the zero eigenvalue \ $\lambda_0=0$. \ To proceed
further, an infinite subtraction is required. It is not really
difficult to understand the origin of this infinity. The
calculation of \ $\mathcal{Z}_{(\mathcal{M}, g)}$ \ takes into
account the contributions of geodesic loops of any length. Each
geometrically distinct geodesic loop \ $C$ \ is counted infinite
times, since even upon traversing it as many times as we want, it
still remains a geodesic loop (although non-minimal). The measure
with which each \ $C$ \  is weighted in \
$\mathcal{Z}_{(\mathcal{M}, g)}$ \ is not sufficient to outweigh
the contributions of all its multiples, so the total contribution
of \ $C$ \ to \ $\mathcal{Z}_{(\mathcal{M}, g)}$ \ is infinite.
The way around such a divergence should now be clear: For a
general compact \ $\mathcal{M}$, \ its diameter \
$diam(\mathcal{M})$ \ which is defined as
\begin{equation}
 diam(\mathcal{M}) \ = \ \mathrm{sup}
 \{ d(p,q), \ \ p,q \in \mathcal{M} \}
\end{equation}
is finite, with \ $d(p,q)$ \ indicating the distance of \
$p,q\in\mathcal{M}$ \ calculated with respect to \ $g$. \ As it
was pointed out above, the classical/non-thermal vacua of interest
have everywhere non-negative sectional curvature. If, moreover,
this curvature is everywhere bounded from below, away from zero,
by \ $\delta>0$ \ then according to Myers' theorem [25] \
$diam\mathcal{M} \leq \frac{\pi}{\sqrt{\delta}}$. \ The curvature
condition of Myers' theorem is satisfied by spheres, by projective
spaces and by the Aloff-Wallach manifolds. Even if \ $\mathcal{M}$
\ does not fulfill the requirements of Myers' theorem though,
there is a weaker estimate for \ $diam\mathcal{M}$. \ If the
Ricci, instead of the sectional, curvature is uniformly bounded
from below, away from zero, by \ $\rho>0$ \ then Cheng's theorem
[25] states that
\begin{equation}
   diam\mathcal{M} \ \leq \ \pi
   \sqrt{\frac{\dim\mathcal{M}-1}{\rho}}
\end{equation}
 Myers' and Cheng's theorems are useful because they provide
upper bounds to a global property of the vacuum, namely its
diameter, which is used in an essential way in the rest of the
argument. It is also important, from a practical standpoint, that
the conditions of these theorems can be checked through local
computations which can be easily performed if one knows the metric
of the vacuum \ $g$. \ The length of the longest possible
piecewise-smooth minimal geodesic loop \ $C$ \ that describes the
evolution of the order parameter/Goldstone field \ $\phi$ \ on \
$\mathcal{M}$ \ during the consecutive bubble collisions should
depend on \ $diam\mathcal{M}$. \ Let such a dependence be denoted
by \ $\alpha (diam\mathcal{M})$. \ It is very difficult to find an
exact expression for \ $\alpha$ \ for a generic \ $\mathcal{M}$. \
For the apparently simpler case of \ $C$ \ being a closed
geodesic, it is not even clear that such an upper bound exists.
However piecewise smooth geodesic loops are less rigid objects
than closed geodesics, exactly because of the almost certain lack
of the differentiability of the former at the points \
$\phi_i\in\mathcal{M}, \  i=0,1,2, \ldots$  \ which are the values
that the order parameter attains in the colliding bubbles. With
such an upper bound on \ $diam(\mathcal{M})$, \ only the piecewise
smooth geodesic loops \ $C$ \ with maximum length \
$\alpha(diam\mathcal{M})$ \ will contribute to the integrated
partition function \ $\mathcal{Z}_{(\mathcal{M},g)}$. \ In other
words, \ $\alpha(diam\mathcal{M})$ \ provides a natural infrared
cutoff for the allowed lengths of geodesic loops \ $C$ \
contributing to  \ $\mathcal{Z}_{(\mathcal{M},g)}$. \ On such,
``physical grounds", (22) reduces to the estimate
\begin{equation}
  \mathcal{Z}_{(\mathcal{M},g)} \ \sim \ \int\limits_0^{\alpha(diam\mathcal{M})} \
         Z_{(\mathcal{M},g)}(t) \ dt
\end{equation}
Then issues of subtractions of infinity from \
$\mathcal{Z}_{(\mathcal{M},g)}$ \ that arose from the existence of
long loops on \ $\mathcal{M}$ \ no longer persist. Substituting
(7) into (28), we get
\begin{equation}
 \mathcal{Z}_{(\mathcal{M},g)} \ \sim \
 \frac{1}{Vol_g(M)} \ \sum_{i=0}^{\infty} \
 \frac{1-e^{-\lambda_i\alpha(diam\mathcal{M})}}{\lambda_i}
\end{equation}
In this equation we immediately see that the contribution of the
zero eigenvalue \ $\lambda_0 = 0$ \ to the sum is \
$\alpha(diam\mathcal{M})$, \ which is finite, a fact which
justifies some of the comments preceding (28). Whether such a
series converges, or is just a formal expression whose infinities
should be dealt with further, depends on \
$\mathrm{Spec}\mathcal{M}$. \ To illustrate this point, consider \
$\mathcal{M}_1 = \mathbb{S}^1$. \ Recalling that \ $\Gamma(1) = 1$
\ and \ $\sum_{k=1}^{\infty}\frac{1}{k^2} = \frac{\pi^2}{6}$, \ we
find
\begin{equation}
   \mathcal{Z}_{(\mathbb{S}^1, g)} \ =  \ \frac{\pi}{12}
\end{equation}
By contrast, let \ $\mathcal{M}_2 = \mathbb{S}^2$. \ Then the
general term of the series is
\begin{equation}
     \frac{2k+1}{k(k+1)}
\end{equation}
from which we can show, using the comparison test, that the series
diverges. For both of the above examples we disregarded the
(infinite) contribution of the zero eigenvalue, as is usually
done in such calculations. \\

To determine the density of topological strings, a set of great
importance is the set of the non-contractible piecewise smooth
geodesic loops \ $\mathcal{L}^\prime\mathcal{M}$, \ which is a
subset of the loop space \ $\mathcal{LM}$. \ To get an estimate
for the number of string-like defects, we have to calculate the
contributions of all such non-contractible piecewise-smooth
geodesic loops and compare these to \
$\mathcal{Z}_{(\mathcal{M},g)}$, \ namely  we want to determine
\begin{equation}
    d_{str} \ = \
    \frac{1}{\mathcal{Z}_{(\mathcal{M},g)}} \
    {\int\limits_{\mathcal{L}^\prime\mathcal{M}}
    Z_{(\mathcal{M},g)}}(t) \ dt
\end{equation}
The expression (32) provides an estimate rather than an exact
prediction for the number density of topological strings produced
in a phase transition. So, we do really expect deviations between
the result of (32) and the experimentally measured or numerically
computed value of the density of topological strings. To highlight
this point we notice that for \ $\mathcal{M} = \mathbb{S}^n, \
n>1$ \ the density is \ $d_{str}=0$ \ as expected due to that \
$\pi_1(\mathbb{S}^n) = 0, \ n>1$. \ For \ $\mathcal{M} =
\mathbb{S}^1$ \ we have
\begin{equation}
 d_{str} \ = \ \frac{12}{\pi^2} \ \sum_{k=1}^{\infty}
 \frac{e^{-k^2\pi}}{k^2} \ = \ 0.062
\end{equation}
This prediction is four times smaller than \ $d_{str} = 0.25$ \
predicted for kinks [27],[28] (in 2 spatial dimensions), and
almost thirteen times smaller than \ $d_{str} = 0.88$ \ predicted
for strings in 3 spatial dimensions [19]. Such discrepancies are
due to the following: the domain of integration of (32) should be
the set of all piecewise smooth loops rather than  the set of
geodesic loops of \ $\mathcal{M}$. \ In reaching (33) we
integrated only over the closed geodesic loops. We missed
therefore a large contribution to \ $d_{str}$ \ coming from the
piecewise-smooth geodesic loops. In general, the number of
piecewise smooth loops over which we have to integrate in (32) is
far higher than the number of geodesic loops, for any vacuum \
$\mathcal{M}$. \ Also, such number of piecewise smooth geodesic
loops should increases at a far higher rate than the number of
geodesic loops, as the length of \ $C$ \ increases. Therefore \
$d_{str}$ \ as was calculated in (32) provides a rough estimate
about the density of topological strings rather than an actual
prediction which can be directly compared with experimental data.
Such numbers will be reasonably close for simple vacua \
$\mathcal{M}$ \ with relatively small diameters and simple
topology. As the metric and topological properties of \
$\mathcal{M}$ \ get more complicated however, we would expect
increasingly larger deviations between the predictions
of (32) and the experimental data.\\

It is probably worth mentioning, at this point, that the dimension
of space-time on which the model with vacuum \ $\mathcal{M}$ \ is
defined is incorporated in the determination of \ $\mathcal{M}$ \
itself. To find \ $\mathcal{M}$ \ one has to carry out
integrations whose value, or even finiteness, depends crucially on
the space-time dimension. To carry out such a calculation  at
non-zero temperature one has to compute the effective potential
[29] which takes into account the contributions of the thermal
fluctuations to the value of the classical potential. The
singularity structure of the effective potential of a model
depends strongly on the space-time dimension [29], a dependence
which is then inherited to \ $\mathcal{M}$. \ Therefore the
dimension of the space-time on which the model is defined is used
in our considerations only very indirectly. Certainly none of our
arguments rely on it. We naively expect that an increase of the
spatial dimension on which the model is defined will result in
large density of defects. We can see this, heuristically, by
noticing that in higher dimensions there are more bubbles
available that can collide with any given bubble, so such an
increase should be reflected by an increase in the density of the
formed defects. Whether this argument can be formally justified or
to what extent it is correct could be the subject of another
investigation.\\

                                 \vspace{3mm}


                         \centerline{\sc References}

                                 \vspace{5mm}

\setlength{\baselineskip}{18pt} \noindent\rm
1. T.W.B. Kibble, \ \emph{J. Phys.} \bf{A9}, \rm 1387, (1976)\\
2. A. Vilenkin, A.P.S. Shellard, \ \emph{Cosmic Strings and
   Other Topological Defects}, \\
   \hspace*{4mm}  Camb. Univ. Press (1994)\\
3. A. Rajantie, \ \emph{Int. J. Mod. Phys.} {\bf A 17}, 1, (2002)\\
4. A.M. Srivastava, \ \emph{Phys. Rev.} \bf{D 45}, \rm  R3304, (1992)\\
5. A.M. Srivastava, \ \emph{Phys. Rev.} \bf{D 46}, \rm  1353, (1992)\\
6. N. Kalogeropoulos, \ \emph{Int. J. Mod. Phys.} {\bf A 21}, 1493, (2006) \\
7. M. Sakellariadou, \ \emph{``The Revival of Cosmic Strings"} in
   ``Proceedings of Pomeranian\\
   \hspace*{4mm} Workshop in Fundamental Cosmology", Poland (2005),
       {\sf arXiv:hep-th/0510227}\\
8.  W.H. Zurek, \ \emph{Act. Phys. Pol.} {\bf B 24}, 1301, (1993)\\
9.  W.H. Zurek, \ \emph{Nature} {\bf 317}, 505, (1985)\\
10.  W.H. Zurek, \ \emph{Phys. Rep.} {\bf 276}, 177, (1996)\\
11. L.M.A. Bettencourt, N.D. Antunes, W.H. Zurek
      \ \emph{Phys. Rev. D} {\bf 62}, 065005, (2000)\\
12. I. Chuang, R. Durrer, N.Turok, B. Yurke, \
                                   \emph{Science} {\bf 251}, 1336, (1991)\\
13. M.J. Bowick, L. Chandar, E.A. Schiff, A.M. Srivastava, \
                           \emph{Science} {\bf 263}, 943, (1994)\\
14. S. Digal, R. Ray, A.M. Srivastava, \ \emph{Phys. Rev.
    Lett.} {\bf 83}, 5030, (1999)\\
15. P.C. Hendry et al., \ \emph{J. Low. Temp. Phys.} {\bf 93}, 1059, (1993)\\
16. C. Bauerle et al., \ \emph{Nature} {\bf 382}, 332, (1996)\\
17. V.M.H. Ruutu et al., \ \emph{Nature} {\bf 382}, 334, (1996)\\
18. M.E. Dodd et al., \ \emph{Phys, Rev. Lett.} {\bf 81}, 3703, (1998)\\
19. T. Vachaspati, A. Vilenkin, \ \emph{Phys. Rev.} {\bf D 30}, 2036, (1984)\\
20. P. Laguna, W.H. Zurek, \ \emph{Phys. Rev. Lett.} {\bf 78}, 2519, (1997)\\
21. N.D. Antunes, L.M. Bettencourt, W.H. Zurek, \
                                     \emph{Phys. Rev. Lett.} {\bf 82}, 2824, (1999)\\
22. S. Digal, A.M. Srivastava, \ \emph{Phys. Rev. Lett.} {\bf 76}, 583, (1996)\\
23. A. Ferrera, \ \emph{Phys. Rev.} {\bf D 59}, 123503, (1999)\\
24. M. Craioveanu, M. Puta, T.M. Rassias, \ \emph{Old and New Aspects in Spectral Geometry}, \\
    \hspace*{5.7mm}  Kluwer Acad. Publ., \ Dordrecht (2001)\\
25. T. Sakai, \ \emph{Riemannian Geometry}, \ Amer. Math. Soc., \
    Providence (1996)\\
26. A.L. Besse, \ \emph{Einstein manifolds}, \  Springer-Verlag, \ Berlin (1987)\\
27. T. Prokopec, \ \emph{Phys. Lett.} {\bf B 262}, 215, (1991)\\
28. R. Leese, T. Prokopec, \ \emph{Phys. Rev.} {\bf D 44}, 3749, (1991)\\
29.  S. Weinberg, \ \emph{The Quantum Theory of Fields,  \ Volume II}, \ \ Camb. Univ. Press,\\
    \hspace*{5.7mm} Cambridge (1996)\\

                                 \vfill

\end{document}